\documentclass[12pt]{article}
\usepackage{epsfig,epsf,amssymb,latexsym}	
\usepackage{color}
\usepackage{multicol}

\begin{document}

\title{\bf A New Perspective to Cosmic Evolution and Vacuum Selection on a Superspace}
\author{Edward Tetteh-Lartey\thanks{lartey@fnal.gov} \\ 
Department of Physics, Texas $A\&M$ University, College Station, TX 77845, 
USA
}

\date{\today}

\maketitle

%\begin{abstract}
\small{
I conjecture that a flat 10D compact Universe emerged out of nothing\footnote{By ``nothing'' I mean no notion of space and time} with all it's symmetries intact: Poincare invariance, conformal invariance and supersymmetry, it's massless moduli fields, and a zero energy density. Its subsequent evolution and spontaneous breaking of these symmetries or by quantum fluctuations leads to topological or geometrical defects on this Superspace which are the fields we observe (or haven't yet observed) propagating in this spacetime. By the law of naturalness it will restore all these broken symmetries. In recent years, the cosmological constant problem has metamorphosed to a new problem of finding a selection mechanism that selects our vacuum out of a string theory landscape or resign to anthropic reasoning.  I provide a new perspective for solving this problem using a multi-step approach based on the wavefunction, inflation and conformal symmetry.
}
%\end{abstract}
%\vspace{2.0in}

\begin{multicols}{2}
\section{Introduction}
\paragraph{}
In recent years the nagging issue of the cosmological constant i.e; why it is so small in contradiction to our quantum field theory predictions, has led to a new problem in physics- finding a vacuum selection mechanism on a vast string theory landscape$~\cite{slandscape}$ that selects our standard model vacuum.  
In addition, we are also confronted with the problem of what observables to define on a dS spacetime, and embedding such time dependent background in perturbative string theory.

In the standard model (gauge theory) there are several observables we can calculate e.g. S-matrix, gauge correlator functions etc; which is done to high accuracy.
The question in string theory is what observables do we need to compute. In string theory the only well defined observable is the S-Matrix, and even only well defined in flat Minkowski spacetime. In AdS spacetimes, it is calculated by using the AdS/CFT duality relation as demonstrated in an AdS Schwarzschild black hole. The boundary correlators are matched to the S-Matrix and the issue of information loss is avoided. In dS spacetime there is an issue of a cosmic horizon. 
Even if we are able to calculate the S-Matrix by using a dS/CFT duality relation as in the AdS case or other, it will be unobservable since the spacetime is expanding so fast it is not possible for observers to observe it no matter how long they wait and they will finally be thermalised by the horizon. Thus the observables in a dS spacetime are called meta-observables\footnote{Meta-observable implies they can be calculated but not observed$~\cite{witten}$}

Due to these severe problems we may turn to non-critical string as I have addressed in$~\cite{etl}$ . Here I assume that the cosmological constant is evolving to a zero value in time due to a decrease in the central charge deficit on the world sheet of the strings. The direction of flow is irreversible, and is directed towards a fixed conformal point with lesser value of the central charge deficit. The implication is that spacetime will eventually be a flat Minkowski spacetime with infinite Hilbert states and hence a well defined S-Matrix.
This is called Q-cosmology. This helps to resolve the issue of S-Matrix, and the vacuum selection problem on the 
string landscape as I have addressed in$~\cite{etl}$. The issue now is whether it can make testable predictions in reasonable agreement with the WMAP data.  

Putting the above issues aside, having a fundamental theory like  string/M theory
explaining the dynamics of the universe is not enough if we should assume that the universe is a quantum mechanical system\footnote{This is a valid
assumption since we have no evidence that the phenomenon we do see on this large scale cannot be described in quantum mechanical terms and explained by quantum mechanical laws, in which case a valid question to ask is what is the quantum state of our Universe}.  If the Universe has a quantum state then a law describing its quantum state must be part of a final theory. The final theory will then be made of two parts:
1) A universal fundamental dynamic law e.g string/M theory
2) A law for the quantum state of the universe e.g Hartle-Hawking's ``No boundary proposal''

The problem of selecting the standard model vacuum on the vast string theory landscape is a very complicated problem. One has to worry about the different possible geometries, choices of Calabi-Yau threefolds, choices of fluxes on the Calabi-Yau manifold and choices of brane configurations which determine the gauge theory. Thus we have a vast landscape of vacua to select from. These are all possible string theory solutions. 

I conjecture that a Universe (Superspace) was born out of {\it nothing} as a flat 10 dimensional compact spacetime with no vacuum energy and in a full supersymmetric, Lorentz invariant, and conformal invariant state, with its massless moduli fields parameterizing the compact dimensions, and satisfying Einstein's vacuum and the Wheeler-DeWitt equations. It's wavefunction given by the Hawking-Hartle wavefunction peaking at zero energy density.
 Thus the universe initially born is non-inflationary but can expand infinitely.
In this note I suggest a new perspective- using multi-step approach to select our vacuum using the wavefunction, the inflationary potentials, and conformal symmetry. I discuss this in section 2.
I conclude in section 3 and highlight on some remaining issues to be tackled.

\section{Multi-step approach}
\paragraph{}
The multi-step approach is as follows:
\begin{itemize}
\item Use the wavefunction of the Universe to select the most probable geometry, thus reducing geometrical landscape. 
\item Use fluxes and branes to generate a potential and stabilize the extra dimensions, giving masses to moduli as well.
\item Use large volume compactification to reduce flux landscape.
\item Use inflation to reduce the landscape of the potentials
\item Use conformal symmetry to reduce the cosmological constant landscape
\item Find a selection mechanism for the standard model gauge theory
\end{itemize}

\subsection{Using the wavefunction of the Universe}
\paragraph{}

The main motivation behind quantum cosmology is a consistent explanation for the origin of the our Universe. The most appealing explanation is the {\it spontaneous creation from nothing}$~\cite{qcos}$. In this picture of the origin of the universe, the Universe is a quantum mechanical system with zero size. There is a potential barrier that it may tunnel with a well-defined, non-zero probability. If the Universe actually tunnels, it emerges to the right of the barrier with a definite size. 
The cosmological wavefunction can be used to calculate the probability distribution for the initial configuration of nucleating universes\footnote{This equation is analogous to the Schrodinger in ordinary quantum mechanics. To solve this equation, one has to specify some boundary conditions for $\Psi$. In quantum mechanics the boundary conditions are determined by the physical set-up external to the system. But since there is nothing external to the universe, it appears that boundary conditions for the wavefunction of the universe should be postulated as an independent physical law.}

The wavefunction $\Psi$ satisfies the Wheeler-DeWitt equation:
\begin{equation}
H\Psi = 0 \label{dwitt}
\end{equation} 
 Solution of this equation requires specification of boundary conditions. There are current three proposals for this: a) The Hartle-Hawking wavefunction, ``No boundary proposal'' b) The Linde$~\cite{linde}$ and Vilenkin$~\cite{vilenkin}$ tunneling wavefunction.
These wavefunctions could be used in describing closed Universes, an infinite topological trivial flat and open Universes, or flat and open Universe with nontrivial topology$~\cite{comp1,comp2}$.
 
Application of these proposals for simple models give different predictions for the initial evolution of the Universe. One of these predictions is the initial energy of the Universe right after its nucleation. The tunneling wavefunction predicts that the Universe must nucleate with the largest possible vacuum energy whereas the no-boundary wavefunction predicts just the opposite$~\cite{hawv}$.   
Also as I pointed out in my work$~\cite{etl}$, both wavefunctions have problems of their own. The tunneling wavefunction is not bounded from above and 
and seems to be in conflict 
with WMAP results on the value of the Hubble parameter
Also the large vacuum energy is inconsistent with a low value of the cosmological constant we observe. The Hartle-Hawking's wavefunction is also not bounded from below,  hence not normalizable and also inconsistent with inflation. But in my multi-step approach, since I will generate the vacuum energy after the Universe has nucleated this is not an issue and I will 
assume
the Hartle-Hawking wavefunction given by $\Psi_{HH} \sim e^{-S_{E}} = \exp\left(\frac{3\pi}{2G_{N}\rho_{v}}\right)$, and a probability distribution peaking at zero energy density ($\rho_{v}=0$). 

Thus a flat 10D compact non-inflationary Universe with a zero vacuum energy is highly favored to nucleate from {\it nothing}. This spacetime is supersymmetric, Poincare invariant, and conformal invariant.

\subsection{Use the fluxes and branes}
\paragraph{}
After selecting the most probably energy and geometry using the wavefunction, the next step is to stabilize six extra dimensions for consistency with our observations of four noncompact dimensions and also give masses to the moduli. The moduli that come with this 10D spacetime have flat directions\footnote{Even if they have potentials, these potentials have runaway behavior and cannot stabilize the moduli} and thus pose a problem. But we get lucky here-
a little while or long after its birth, quantum fluctuations or spontaneous breaking of a symmetry could lead to structures or defects such as branes, fluxes, strings etc in this spacetime.  In the case of 11D M-Theory, these defects could be a result of the collapse of the 11th dimension in order to restore Lorentz invariance. As the Superspace expands the quantun flunctuations and hence these defects decreases. 

The wrappings of the fluxes and branes on six of the ten dimensions as demonstrated by the KKLT Mechanism$~\cite{kklt}$ generates a potential (vacuum energy) stabilizing the moduli. I must point out that not all regions of this Superspace may necessarily have these defects or have their moduli stabilized\footnote{Note the importance of moduli stabilization in our patch- without it the extra dimensions will be noncompact, inconsistent with observation and possible fifth long range force violating the equivalence principle}. This potential generation and moduli stabilization may only have occurred in a small patch of the superspace leaving the vast section of the Superspace with infinite 10 dimensions.   
 
\subsection{Large volume flux compactification}

As argued in$~\cite{con, conthesis}$ in D3/D7 IIB orientifold compactification, if we wish to study the scalar potential across the full range of flux choices and moduli values, it is essential to include perturbative corrections to the $K\ddot{a}hler$ potential {\it K}. The full $K\ddot{a}hler$ potential {\it K}, and superpotential {\it W} are given by:

\begin{equation}
{\it K} = {\it K_{0}} + \it K_{p} + \it K_{np} \approx \it K_{0} + \it A \label{kah}
\end{equation}

\begin{equation}
{\it W} = {\it W_{0}} + \it W_{np} \approx \it W_{0} + \it B \label{kah2}
\end{equation}

where {\it A} represents the leading perturbative correction to {\it K}
and {\it B} the leading nonperturbative correction to {\it W}. ${\it K_{0}}$ and ${\it W_{0}}$ are the tree level $K\ddot{a}hler$ potential and superpotential respectively.

The F-term scalar potential is given by:

\begin{equation}
V_{F} = e^{\it K}[\it K^{i\overline{k}}\it D_{i}W\it D_{\overline{k}}\overline{\it W} - 3|{\it W}|^{2}] \label{sca} 
\end{equation}

which can be expanded in powers of {\it A} and {\it B}:
\begin{equation}
V_{F} = V_{0} + V_{\it A} + V_{\it B} + .... \label{sca2}
\end{equation}

where
\[
 V_{0}\sim \it W_{0}^{2}, \ V_{\it A}\sim \it AW_{0}^{2}, \ V_{\it B}\sim \it B^{2} +  \it W_{0}B.
\]

Perturbative corrections to the scalar potential {\it (0(A) terms)}, are negligible depending on the value of the tree level superpotential. Perturbative corrections are not necessary if ${\it W_{0}}$ is extremely small. As pointed in$~\cite{con}$, the limit to which the perturbative corrections could be neglected is unnatural and should therefore be included in most cases.
These perturbative corrections to the $K\ddot{a}hler$ potential\footnote{The leading corrections to the $K\ddot{a}hler$ potential were computed in$~\cite{cor}$ and arise from the ten-dimensional {\it O($\alpha^{\prime 3}$)} ${\it R^{4}}$ term. The $\alpha^{\prime}$ expansion is an expansion in inverse volume and thus at large volume can be controlled.} in addition to the nonperturbative contributions to the superpotential generically gives rise to a large volume non-supersymmetric AdS vacuum, differing from the simplest KKLT scenario in which the AdS minimum is found to be supersymmetric.  

In this large volume limit, supersymmetry is broken by the $K\ddot{a}hler$ moduli only, and
 the gravitino mass is independent of choices of flux on the Calabi-Yau manifold, thus do not scan from vacuum to vacuum  as the fluxes are tuned but peaked at a particular Calabi-Yau.
The universality of the gravitino mass across the space of flux choices, could indicate universality or near universality across the space of Calabi-Yau manifolds. 

This large volume limit brings us closer to lower energy scale physics. 
For a stabilized Calabi-Yau threefold, the gravitino mass and string scale are given by:

\begin{equation}
m_{3/2} \sim \frac{M_{p}}{\it V}, \ m_{s}\sim \frac{M_{p}}{\sqrt{\it V}} \label{grav}
\end{equation}

where {\it V} is the dimensionless volume- the physical volume is ${\it Vl_{s}^{6}} \equiv {\it V(2\pi\sqrt{\alpha^{\prime}})^{6}}$. Thus a compactification volume of $10^{15}{\it l_{s}^{6}}$, corresponding to a string scale $m_{s}\sim 10^{11}GeV$, can generate the weak hierarchy through TeV-scale supersymmetry$~\cite{tscale}$
 
\subsection{Use inflationary potentials}
\paragraph{}
At this stage we can have the moduli stabilized but then we encounter several possible configurations of branes and choices of fluxes generating potentials that could stabilized the six compact dimensions. Here the next selection mechanism is to use inflationary potentials. 

The difficulty of obtaining slow-roll inflation acts as a selection mechanism on the different possible configurations of branes and fluxes wrapping the extra dimensions. Regions with potentials not providing the right conditions for slow-roll inflation will be quickly dominated by regions with inflationary potentials\footnote{Note that matter fields only come into existence on exit from slow-roll inflation when the inflaton oscillates near the minimum of its potential. This is called the reheating phase}.
The dominant potential(s) may then undergo slow roll eternal inflation (SREI)  
 dividing this patch of Superspace into many exponentially large domains corresponding to different metastable vacuum states, and forming a huge inflationary multiverse. This gives a landscape of cosmological constants$~\cite{slandscape}$. 

From the Superspace point of view, the branes with either flat or curved geometry are 
defects which breaks Lorentz invariance in that region of spacetime and needs to be corrected. Inflation helps in smoothing this out making it flat.
These branes or defects from which our Universe emerges are like black holes propagating in this spacetime with a comic horizon which are the edges pushing against the Superspace.
 
Observers on the brane experience a local cosmic horizons.  There may be several branes:- a hidden sector where the vacuum energy was generated leading to breaking of supersymmetry and the standard model sector on which we live.

\subsection{Use conformal symmetry}
 \paragraph{}

One must remember that a very crucial symmetry has been broken by the strings propagating on this landscape due to the vacuum energy. The world-sheet of the strings are not conformal. As I have explained in$~\cite{etl}$ a temporal restoration of conformal symmetry is done by Liouville dressing with a permanent restoration by the renormalization group flow, reducing the central charge deficit to zero conformal fixed point.

This expansion and departure from criticality from a conformal point is taking place in a linear dilaton background. Each of this regions with different values of the cosmological constant is a departure from criticality with a tendency to restore conformal symmetry. Regions with the largest vacuum energies i.e largest central charge deficit may have the strongest tendency. Over a period of time all regions will merge into one and move towards the zero conformal point.  
We are at the point where this renornamisation group flow has a central charge deficit (vacuum energy) condusive for life and we have emerged as a result.

Some questions worth asking are:
\begin{itemize}
\item A linear dilaton background allows for any number of spacetime dimensions for conformal invariance so why 10D. The simple answer is that, it is the tendency of nature to have the maximum symmetry. 10D gives Lorentz invariance though I must state that the presence of vacuum energy will cause Lorentz invariance to be violated. 
\item Why not a constant dilaton background where 10D will be the requirement for conformal invariance- The answer could come from M-Theory. The dilaton is a field resulting from the collapse of the 11th dimension. It could however be stabilized as done in flux compactifications. But here we must assume that it is stabilized at a value equal or less than zero value for perturbative theory to be valid. 
\item Why only six dimensions are stabilized- I do not yet have an explanation for this, but it could be due to the topological or geometrical structure of the fluxes.  A good answer to this could help explain why we live on D3-brane and not any other D brane of different dimensionality. 
Also, it is possible that the four fundamental forces we observe have a relation to the unstabilized moduli fields in the four noncompact dimensions. This needs to be investigated.
\end{itemize}
The linear dilaton decreasing towards the conformal point due to the decreasing central charge deficit shows that a move from a strong coupling regime to a weak coupling at future times, $t\rightarrow \infty$. This makes perturbation expansions valid.   
In the strong coupling limit i.e at very early times with a 
largest value of the dilaton's expectation value, an eleventh dimension could fluctuate into existence but this is transient and re-collapses back into the Superspace due to a need to restore Lorentz symmetry.
This dimension could collapse into fields like graviton, gauge fields, scalar fields etc. propagating in this spacetime.
  
\subsection{Select the gauge theory}
\paragraph{}
At this stage we have reduce the vacau substantially. We must now devise a mechanism that selects our standard model gauge group.
The standard model gauge theory of $SU(3)\times SU(2)\times U(1)$ is believed to be a result of stacks of intersecting Dp-branes e.g, D6 branes wrapping 3-cycles on the Calabi-Yau manifold. $N_{c}$ coincident D-branes give an $SU(N_{c})$ gauge theory. At another location on the manifold, another stacks of branes generates a potential leading to supersymmetry breaking. This is called the hidden sector. One must understand the mechanism selecting this configuration out of all other possibilities. 
\begin{center}
\includegraphics[width=4.0in]{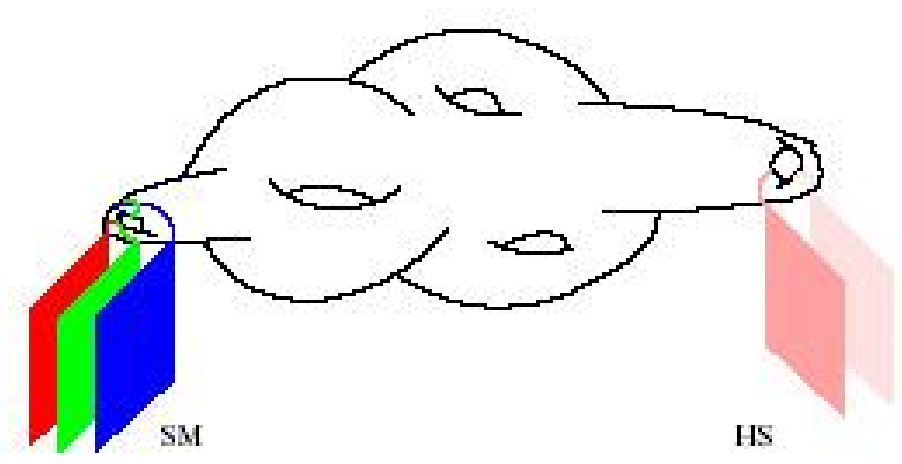} 
{\it Fig. 1. A Calabi space with intersecting standard model and hidden D6-branes$~\cite{lust}$}
\end{center} 
I don't have much to say on this at the moment but further work in this area will be useful.

\section{Conclusion}
\paragraph{}
Symmetry principles have always been very useful in resolving issues in physics. Solving the cosmological constant problem from some symmetry principles such as supersymmetry have not been very helpful. In this note I suggest that we have not yet exhausted this use of symmetry and thus conformal symmetry provides a solution. This is naturalness\footnote{Naturalness as argued by 't Hooft states that if a quantity is much smaller than expected from dimensional analysis, this should be because the theory becomes more symmetric in the limit that the quantity tends to zero} occurring through conformal symmetry.
 
We are living on a brane inside a giant Superspace. The branes are spacetime defects arising from a broken symmetry, quantum fluctuations or collapse of the 11th dimension of M-theory.  
Our patch of this Superspace is on its way to restoring conformal symmetry permanently and has reach a value along the renormalization group flow which is suitable for life.  

A dS spacetime is an intermediary stage in the quantum evolution of the Universe. Eventually, spacetime will have a flat Minkowski geometry, where all its symmetries- supersymmetry, conformal, and  Poincare invariance will be restored. We are living at the very edge of this full restoration in the patch of this giant Superspace. This period provides a suitable condition for life and thus we have popped into existence.
  
If by Liouville dressing we could get temporal conformal invariance, then the dS/CFT correspondence could be possible if we could identify and match the correlation functions on the gravity side and the field theory side. 

It is possible that the four fundamental forces we experience have a relation to the unstabilized moduli in our four noncompact dimensions. Further work is needed in this area as well as finding a selection mechanism for the standard model gauge group.

\end{multicols}


\begin{thebibliography}{99}
\bibitem{kklt}Shamit Kachru, Renata Kallosh, Andrei Linde and Sandip P. Trevedi, ``de Sitter Vacua in String Theory'', hep-th/0301240.
\bibitem{slandscape}L. Susskind, ``The Anthropic Landscape of String Theory,'' arXiv:hep-th/0302219
\bibitem{witten} E. Witten, ``Quantum Gravity in De Sitter Space'', hep-th/0106109

\bibitem{qcos}J. B. Hartle and S. W. Hawkings, Phys. Rev D. {\bf 28}, 2960 (1983).
\bibitem{linde} A. D. Linde, Sov. Phys. JETP 60, 211 (1984).
\bibitem{vilenkin}A. Vilenkin, Phys. Rev. D27, 2848 (1983).
\bibitem{hawv}A. Vilenkin in Cambridge 2002, The future of theoretical physics and cosmology, eds. G. W. Gibbons, E. P. S Shellard and S. J. Rankin (Cambridge University Press, Cambridge 2003), 649-666.
\bibitem{comp1} A. Friedmann, ``The Universe as Space and Time,'' Leningrad, Akademiya (1923); Phys. {\bf 21}, 326 (1924).
\bibitem{comp2}Y. B. Zeldovich and A. A. Starobinsky, ``Quantum Creation of a Universe in a Nontrivial Topology,'' Sov. Astron. Lett. {\bf 10}, 135 (1984).
\bibitem{etl}E. Tetteh-Lartey, ``Vacuum Selection on the String Landscape,'Phy. Rev D${\bf 75}$, 106005, (2007), hep-th/0703160.
\bibitem{con}V. Balasubramanian, P. Berglund, J. P. Conlon, and F. Quevedo, ``Systematics of Moduli Stabilization in Calabi-Yau Flux Compactification,'' arXiv:hep-th/0502058 
\bibitem{cor}K. Becker, M. Becker, M. Haack, and J. Louis, ``Supersymmetry breaking and $\alpha^{\prime}$-corrections to flux induced potentials,'' JHEP {\bf 06} (2002) 060, hep-th/0204254.
\bibitem{conthesis} J. Conlon, ``Moduli Stabilization and Applications in IIB Theory,''hep-th/0611039.
\bibitem{tscale}C.P. Burgess, L. E. Ibanez and F. Quevado, Phys. Lett. B{\bf 447} (1999) 257, arXiv:hep-ph/9810535; K. Benakli, Phys. Rev. D{\bf 60} (1999) 104002, hep-ph/9809582 
\bibitem{lust}D. Lust, ``Intersecting branes-a path to the standard model?,'' Class. Quantum Grav. {\bf 21} (2004)S1399-S1424. 
\end{thebibliography}
\end{document}